\documentclass[
  amsmath,
  amssymb,
  aps,
  twocolumn,floatfix]{revtex4-2}
\PassOptionsToPackage{hyphens}{url}
\PassOptionsToPackage{dvipsnames,svgnames,x11names}{xcolor}
\usepackage{amsmath,amssymb}
\usepackage{iftex}
\ifPDFTeX
  \usepackage[T1]{fontenc}
  \usepackage[utf8]{inputenc}
  \usepackage{textcomp} 
\else 
  \usepackage{unicode-math}
  \defaultfontfeatures{Scale=MatchLowercase}
  \defaultfontfeatures[\rmfamily]{Ligatures=TeX,Scale=1}
\fi
\usepackage{lmodern}
\ifPDFTeX\else  
\fi
\IfFileExists{upquote.sty}{\usepackage{upquote}}{}
\IfFileExists{microtype.sty}{
  \usepackage[]{microtype}
  \UseMicrotypeSet[protrusion]{basicmath} 
}{}
\makeatletter
\@ifundefined{KOMAClassName}{
  \IfFileExists{parskip.sty}{%
    \usepackage{parskip}
  }{
    \setlength{\parindent}{0pt}
    \setlength{\parskip}{6pt plus 2pt minus 1pt}}
}{
  \KOMAoptions{parskip=half}}
\makeatother
\usepackage{xcolor}
\ifx\paragraph\undefined\else
  \let\oldparagraph\paragraph
  \renewcommand{\paragraph}[1]{\oldparagraph{#1}\mbox{}}
\fi
\ifx\subparagraph\undefined\else
  \let\oldsubparagraph\subparagraph
  \renewcommand{\subparagraph}[1]{\oldsubparagraph{#1}\mbox{}}
\fi

\usepackage{booktabs,array}
\usepackage{calc} 
\usepackage{etoolbox}
\IfFileExists{footnotehyper.sty}{\usepackage{footnotehyper}}{\usepackage{footnote}}
\usepackage{graphicx}
\makeatletter
\def\maxwidth{\ifdim\Gin@nat@width>\linewidth\linewidth\else\Gin@nat@width\fi}
\def\maxheight{\ifdim\Gin@nat@height>\textheight\textheight\else\Gin@nat@height\fi}
\makeatother
\setkeys{Gin}{width=\maxwidth,height=\maxheight,keepaspectratio}
\makeatletter
\def\fps@figure{htbp}
\makeatother
\newlength{\cslhangindent}
\newlength{\csllabelwidth}
\newlength{\cslentryspacingunit} 
\newenvironment{CSLReferences}[2] 
 {
  \setlength{\parindent}{0pt}
  \ifodd #1
  \let\oldpar\par
  \def\par{\hangindent=\cslhangindent\oldpar}
  \fi
  \setlength{\parskip}{#2\cslentryspacingunit}
 }%
 {}
\usepackage{calc}

\newcommand{\CSLLeftMargin}[1]{\parbox[t]{\csllabelwidth}{#1}}
\newcommand{\CSLRightInline}[1]{\parbox[t]{\linewidth - \csllabelwidth}{#1}\break}

\usepackage{physics}
\makeatletter
\AtBeginDocument{%
\ifdefined\contentsname
  \renewcommand*\contentsname{Table of contents}
\else
  \newcommand\contentsname{Table of contents}
\fi
\ifdefined\listfigurename
  \renewcommand*\listfigurename{List of Figures}
\else
  \newcommand\listfigurename{List of Figures}
\fi
\ifdefined\listtablename
  \renewcommand*\listtablename{List of Tables}
\else
  \newcommand\listtablename{List of Tables}
\fi
\ifdefined\figurename
  \renewcommand*\figurename{Figure}
\else
  \newcommand\figurename{Figure}
\fi
\ifdefined\tablename
  \renewcommand*\tablename{Table}
\else
  \newcommand\tablename{Table}
\fi
}
\@ifpackageloaded{float}{}{\usepackage{float}}
\floatstyle{ruled}
\@ifundefined{c@chapter}{\newfloat{codelisting}{h}{lop}}{\newfloat{codelisting}{h}{lop}[chapter]}
\floatname{codelisting}{Listing}

\makeatother
\makeatletter
\@ifpackageloaded{caption}{}{\usepackage{caption}}
\@ifpackageloaded{subcaption}{}{\usepackage{subcaption}}
\makeatother
\makeatletter
\@ifpackageloaded{tcolorbox}{}{\usepackage[skins,breakable]{tcolorbox}}
\makeatother
\makeatletter
\@ifundefined{shadecolor}{\definecolor{shadecolor}{rgb}{.97, .97, .97}}
\makeatother
\makeatletter
\ifdefined\Shaded\renewenvironment{Shaded}{\begin{tcolorbox}[breakable, borderline west={3pt}{0pt}{shadecolor}, sharp corners, interior hidden, frame hidden, enhanced, boxrule=0pt]}{\end{tcolorbox}}\fi
\makeatother
\ifLuaTeX
  \usepackage{selnolig}  
\fi
\IfFileExists{bookmark.sty}{\usepackage{bookmark}}{\usepackage{hyperref}}
\IfFileExists{xurl.sty}{\usepackage{xurl}}{} 
\hypersetup{
  pdftitle={Efficient, direct compilation of SU(N) operations into SNAP \& Displacement gates},
  pdfauthor={Joshua Job},
  colorlinks=true,
  linkcolor={blue},
  filecolor={Maroon},
  citecolor={Blue},
  urlcolor={Blue},
  pdfcreator={LaTeX via pandoc}}

\usepackage{mathtools}

\begin{document}

\title{Efficient, direct compilation of SU(N) operations into SNAP \&
Displacement gates}
\author{Joshua Job}
\affiliation{Lockheed Martin Space, Sunnyvale, CA, 94089 and Superconducting Quantum Materials and Systems Center (SQMS), Fermilab}
\date{\today}

\begin{abstract}
  We present a function which connects the parameter of a previously
  published short sequence of selective number-dependent arbitrary phase
  (SNAP) and displacement gates acting on a qudit encoded into the Fock
  states of a superconducting cavity,
  \(V_k(\alpha)=D(\alpha)R_\pi(k)D(-2\alpha)R_\pi(k)D(\alpha)\) to the angle of the Givens
  rotation \(G(\theta)\) on levels \(\ket{k},\ket{k+1}\) that sequence approximates,
  namely \(\alpha=\Phi(\theta) = \frac{\theta}{4\sqrt{k+1}}\). Previous publications left
  the determination of an appropriate \(\alpha\) to numerical optimization at
  compile time. The map \(\Phi\) gives us the ability to compile directly any
  \(d\)-dimensional unitary into a sequence of SNAP and displacement gates
  in \(\mathcal{O}(d^3)\) complex floating point operations with low
  constant prefactor, avoiding the need for numerical optimization.
  Numerical studies demonstrate that the infidelity of the generated gate
  sequence \(V_k\) per Givens rotation \(G\) scales as approximately
  \(\mathcal{O}(\theta^6)\). We find numerically that the error on
  compiled circuits can be made arbitrarily small by breaking each
  rotation into \(m\) \(\theta/m\) rotations, with the full \(d\times d\)
  unitary infidelity scaling as approximately \(\mathcal{O}(m^{-4})\).
  This represents a significant reduction in the computational effort to
  compile qudit unitaries either to SNAP and displacement gates or to
  generate them via direct low-level pulse optimization via optimal
  control.
  \end{abstract}

\maketitle

\hypertarget{introduction}{%
\section{Introduction}\label{introduction}}

The prospect of using high-dimensional qudits for quantum computation
rather than qubits is an enticing possibility. While one may always
represent any finite dimensional model in qubit form, it comes at a
penalty of \(\order{\log{d}}\) overhead in the number of qubits. The
hope is that by being able to directly represent high-dimensional
sub-systems in individual qudits we can create more efficient
constructions for problems in high energy physics, condensed matter, and
other areas.

The design and implementation of gates for such high-dimensional qudits
is not trivial. One can design gates directly to implement useful
operations, such as a quantum Fourier transform (QFT) which either
directly or as an indirect component of quantum phase estimation (QPE)
finds uses in many high energy physics, optimization, and machine
learning applications among others ~{[}4{]}, but compiling that abstract
unitary to a pulse sequence or set of elementary gates on a qudit system
may be computationally intensive.

In this paper, we briefly review existing approaches for constructing
qudit gates, focusing on qudits formed by modes of a superconducting
cavity, and then demonstrate an efficient decomposition scheme into
displacement and selective number-dependent arbitrary phase (SNAP)
gates. While our proposed scheme builds on previous work, in particular
Refs. ~{[}5,6{]} and ~{[}7{]}, we are to our knowledge the first to
demonstrate how to directly synthesize any SU(N) unitary for a qudit
encoded in the Fock states of a superconducting cavity without any form
of numerical optimization at compile time.

\hypertarget{pulse-level-approaches-and-computational-complexity}{%
\section{Pulse level approaches and computational
complexity}\label{pulse-level-approaches-and-computational-complexity}}

In the most general description, quantum control on a system with
Hamiltonian \(H_\theta\) and control Hamiltonian operators denoted by
\(\{H_c\}\), the full system with controls has the time dependent
Hamiltonian:

\[H(t) = H_0 + \sum_c f_c(t) H_c \]

Here, \(f_c(t)\) denote time-dependent amplitudes for each control
operator, which we may parameterize in various ways. Different
parameterizations generally naturally yield different quantum control
algorithms.

For instance, discretizing \(f_c(t)\) as a sequence of piecewise
constant functions yields a state evolution as described by a product of
unitaries in the form:

\[ U(\{f_c\}) = \prod_j \exp{-i (H_0+\sum_c f_c(j\Delta t)H_c) \Delta t)} \]

This is the approach of the GRAPE algorithm ~{[}8{]}. One may instead
represent the control functions as composed of a sum of carrier
frequencies modulated by envelopes parameterized by B-splines and
perform optimization using a St"ormer-Verlet discretized adjoint scheme,
which yields the proposal implemented in the \texttt{Juqbox.jl}
~{[}9--11{]}. There are many other such proposals. Their common goal is
to find a set of parameters such that the evolution induced by \(H(t)\)
results in a unitary ``close'\,' by some metric to one's target gate.

All of these techniques share similar scaling properties. GRAPE requires
multiplying \(T/\Delta t\) \(N\times N\) matrices along with doing an equal
number of matrix exponentiations for a total gate time \(T\) and
timestep \(\Delta t\). Both matrix exponentiation and matrix-matrix
multiplication are (for most practical purposes) \(\order{N^3}\) for an
\(N \times N\) unitary. Typically \(T/\Delta t\gg \order{N}\) for an \(N\)-d
system, since for a constant number of control terms \(H_c\) it requires
\(\order{N^2}\) timesteps in order to reach the \(\order{N^2}\)
parameters in an \(N \times N\) unitary. This means that calculating
optimal controls using GRAPE optimization will tend to require a runtime
as high (based on this parameter counting argument) as \(\order{N^5}\).
This is only the complexity of a single update pass for the algorithm,
and does not include any scaling in the number of optimization
iterations necessary to find a control sequence which produces the
desired gate to high accuracy.

Techniques based on integrating the Schr\"odinger equation
\[\partial_t U(t) = -i H(t)U(t)\] and variations, ala Juqbox, GOAT
~{[}12{]}, and others requires a minimum of \(O(N^2)\) operations for
dense unitaries per time-step of the integration, but much like GRAPE
can be expected to require a number of time steps that scales with
system size along with a number of timesteps to converge which scales
with size as well. For instance in Juqbox the number of timesteps should
grow as \(N^2\), as in Eq. 26 in Ref ~{[}9{]}. This scaling can also be
explained from a parameter counting argument, since the integration over
\(K\) timesteps is in effect described by \(\order{K\abs{\{H_c\}}}\)
parameters, and for a constant number of control parameters
\(\abs{\{H_c\}}\) one needs \(K=\order{N^2}\) to have sufficiently many
parameters to in principle describe any \(N\)-dimensional unitary. Thus,
these methods will typically also have a runtime that scales as
\(\order{N^4}\).

For very large qudits, this scaling may become an issue. For instance,
it takes approximately twenty minutes in the study presented in Table 5
in Ref. ~{[}9{]} to optimize a high fidelity SWAP gate on an
8-dimensional qudit. Under an extremely optimistic assumption of
\(\order{N^3}\) scaling (as perhaps the unitary is very special and only
requires \(\order{N}\) parameters), it would require weeks to optimize
an operation on \(100\) states, and years to do so on \(\order{500}\)
states.

This is likely impractical, since we expect parameters of our qudits to
shift slowly over time, which may require recomputation of the optimal
control sequences periodically. To address this potential issue, we may
turn to a simple universal gate set, such as SNAP and displacement gates
for a superconducting cavity dispersively coupled to a qubit.

\hypertarget{snap-displacement-gates-for-universal-control}{%
\section{SNAP \& Displacement gates for universal
control}\label{snap-displacement-gates-for-universal-control}}

For qudits encoded into the Fock states of one of the modes of a
superconducting cavity, it has been shown that selective
number-dependent arbitrary phase (SNAP) and displacement gates are
computationally universal ~{[}6{]}. A general SNAP gate can be
represented as a simple diagonal operator,
\[ S(\vec{\theta}) = \sum_n \exp{\theta_n} \op{n} \] which applies, as its name
suggests, an arbitrary phase to each Fock state of the oscillator.
Displacement gates coherently pump or remove energy (photons) from the
oscillator \[D(\alpha)=\exp(\alpha\hat{a}^\dagger - \alpha^* \hat{a})\] for arbitrary
complex \(\alpha\).

Because these two gates together are universal, a sufficiently long
sequence of alternating SNAP and displacement gates can generate an
arbitrary approximation of any unitary. Moreover, we can restrict
\(\alpha \in \mathbb{R}\), as one can shift the phase of a complex
\(\alpha = r\exp{i\phi}\) into the adjacent SNAP gates via
\(S(\vec{\theta})D(r)S(-\vec{\theta})\) with \(\theta_n = n\phi\) ~{[}6{]}. From here on
we will assume real \(\alpha\).

To construct a target unitary \(U_t\) from a sequence of SNAP and
displacement gates, there are several potential approaches.

The first is direct numerical optimization via gradient descent over the
full space of SNAP and displacement parameters, by defining
\[U(\{\vec{\alpha}_j\},\vec{\alpha}) = \prod_j S(\vec{\theta}_j)D(\alpha_j)\] and minimizing
a loss function with respect to the parameters of the gates, for example
the infideltiy \(\epsilon = 1-F\) for fidelity \(F\)
\[F = \frac1d\abs{\Tr(U(\theta,\alpha)^\dagger U_t)}\] Typically these
optimizations are performed in a truncated Hilbert space of dimension
\(d+g\) where \(d\) is the dimension of your qudit (the number of
computationally useful levels you intend to address) while \(g\) denotes
some number of guard states, used to ensure accuracy of the truncation.
This sort of large scale brute-force optimization has been used, for
instance to optimize for a quantum fourier transform (QFT) gate in Ref.
~{[}2{]}.

Innovations on this scheme have been proposed, such as first performing
a nested sequence of local optimizations by injecting short sequences of
SNAP and displacement gates inside the an existing string of gates
thereby iteratively growing a high quality initial ansatz for the gate
sequence, and only then performing a global descent pass to fine-tune as
in Ref. ~{[}7{]}. Ref. ~{[}7{]} was also important in pointing out
that in principle the length of the sequence of SNAPs and displacements
can be lower bounded from a parameter counting argument to be
\(\order{d}\), since each SNAP gate can in principle add \(d-1\) free
parameters (as global phases are irrelevant), and a displacement gate
adds one free parameter. Thus a sequence of alternating SNAP and
displacements with \(\order{d}\) of each will have the same number of
free parameters as a \(d \times d\) unitary.

However, these methods suffer from scaling challenges. By their nature
they require, for each gradient step, at minimum performing a forward
pass through the sequence which requires \(d\) full rank matrix-matrix
multiplications, each of practical time \(d^3\), implying the a scaling
of \(\order{d^4}\) for each gradient step, with an unknown number of
steps being required to find a high quality solution.

Another technique, originally introduced in Ref ~{[}6{]}, is to note
that any unitary operation can be decomposed into a sequence of SNAP
gates followed by SO(2), or Givens, rotations between adjacent Fock
states. The method, as presented there, is as follows:

\begin{itemize}
\item
  To build a unitary \(U\) on a subspace, take the inverse
  \[ U^\dagger = \left(\begin{array}{c|c}
  W &  0\\
  \hline
  0 & I
  \end{array}\right)\] where \(W\) is the non-trivial block, acting on
  your qudit, and \(I\) is the identity.
\item
  First apply a SNAP gate to render all terms in the final column of
  \(W\) non-negative.
\item
  Then, apply an SO(2), or Givens, rotation on the top-right
  \(2\times 2\) block of \(W\), call it
\end{itemize}

\(\big(\begin{smallmatrix}W_{1}^{k-1} & W_{1}^{k} \\ W_{2}^{k-1} & W_{2}^{k}\end{smallmatrix}\big)\),
such that \(W_{1}^{k}=0\).

\begin{itemize}
\item
  Repeat as one goes down the final column of \(W\), setting each value
  to 0 until one reaches the final \(2 \times 2\) block. Unitarity
  ensures that the final element in the column is \(1\) and all elements
  in the final row are \(0\).
\item
  Repeat on the remaining non-identity block for each column until one
  has produced the full inverse.
\item
  The sequence of SNAP and Givens rotation operations applied have
  transformed \(U_t^\dagger\) into \(I\), and thus must have formed
  \(U_t\).
\end{itemize}

This procedure is attractive as given a target unitary one can
immediately construct a set of SNAP and Givens rotations which generate
that unitary.

In Ref ~{[}6{]}, it was found by numerical optimization that a Givens
rotation, \(G(\theta)\), between states \(\ket{k}\) and \(\ket{k+1}\) can be
approximately generated by a short sequence \begin{equation}
    V_k(\alpha) = D(\alpha)R_\pi(k)D(-2\alpha)R_\pi(k)D(\alpha)
\end{equation} where \(R_\pi(k)\) is a SNAP gate of the form
\(\sum_{j=0}^{k} exp(i \pi)\op{k}\) for an appropriate choice \(\alpha\).

Unfortunately, the correct choice of \(\alpha\) was left to numerical
optimization for each Givens rotation, and the final gate sequence,
defined now by a single vector of \(\order{d^2}\) \(\alpha\) values along
with the set of \(d\) SNAP gates, was numerically optimized in full to
fine-tune the unitary.

Unfortunately, numerically optimizing \(V_k(\alpha)\) to generate a given
Givens rotation \(G(\theta)\) requires gradient descent and matrix
exponentiation at each step, or equivalently two matrix-matrix
multiplies if one employs a one-time singular value decomposition of
\((a+a^\dagger)\), the generator of the displacement gate. This is,
again, a \(\order{d^3}\) operation, and must be performed
\(\order{d^2}\) times for a total runtime of \(\order{d^5}\).

However, if it were possible to \emph{directly} map from the target
angle of of a Givens rotation \(G(\theta)\) between any two adjacent Fock
states to the displacement magnitude \(\alpha\) in \(V_k(\alpha)\), ie
\(\alpha = \Phi(\theta)\) for some constant run-time function \(\Phi\), one could
eliminate numerical optimization from gate construction altogether and
produce an arbitrary unitary with minimal overhead.

Such a map \(\Phi\) is presented here.

\hypertarget{first-order-expansion}{%
\section{First order expansion}\label{first-order-expansion}}

To begin, let us expand \(V_k(\alpha)\) to first order in \(\alpha\). For
notational ease, we write \(\hat{A}=\hat{a}^\dagger-\hat{a}\) so that
for real \(\alpha\),
\(D(\alpha) = \exp{\alpha(\hat{a}^\dagger-\hat{a}} = \exp{\alpha\hat{A}}\), and we will
write \(R_\pi(k)\) as \(R_k\), and we note that \(R_\pi(k)^2 = I\)

  \begin{multline*}
    V_k(\alpha)= D(\alpha)R_kD(-2\alpha)R_kD(\alpha) \\
      \hspace{2.5em} = (I+\alpha\hat{A})R_k(I-2\alpha\hat{A})R_k(I+\alpha\hat{A}) + \order{\alpha^2} \\
  \hspace{-0.9em}=I+\alpha\hat{A}-2\alpha R_k\hat{A}R_k+\alpha\hat{A} +\order{\alpha^2} \\
    \hspace{-3.5em}=I+2\alpha[\hat{A}-R_k\hat{A}R_k]+\order{\alpha^2} \\
    \hspace{-14.5em}\text{or, in the full notation:} \\
   \hspace{-1.5em} V_k(\alpha)\approx I+2\alpha\big[(\hat{a}^\dagger-\hat{a}) -R_\pi(k)(\hat{a}^\dagger-\hat{a})R_\pi(k)\big]\\
\end{multline*}
 up to \(\order{\alpha^2}\). The action of \(R_\pi(k)\hat{O}R_\pi(k)\) on an
operator \(\hat{O}\) can be written in block form 
\begin{align*}
\hat{O}&=
\begin{pmatrix}
\hat{O}_{0:k}^{0:k} & \hat{O}_{k}^{k+1:d} \\
\hat{O}_{k+1:d}^{0:k} & \hat{O}_{k+1:d}^{k+1:d}
\end{pmatrix} \\
R_\pi(k)\hat{O}R_\pi(k)&=
\begin{pmatrix}
\hat{O}_{0:k}^{0:k} & -\hat{O}_{0:k}^{k+1:d} \\
-\hat{O}_{k+1:d}^{0:k} & \hat{O}_{k+1:d}^{k+1:d}
\end{pmatrix}
\end{align*}

Therefore, writing \(\hat{A}=\hat{a}^\dagger-\hat{a}\) in the same block
form we get that \(V_k(\alpha)\) 
\begin{align*}
    V_k(\alpha) &\approx \begin{pmatrix}I+2\alpha(\hat{A}-\hat{A})_{0:k}^{0:k} & 2\alpha(\hat{A}+\hat{A})_{0:k}^{k+1:d}\\
    2\alpha(\hat{A}+\hat{A})_{k+1:d}^{0:k} & I+2\alpha(\hat{A}-\hat{A})_{k+1:d}^{k+1:d}
    \end{pmatrix} \\
    &= \begin{pmatrix}
    I & 4\alpha\hat{A}_{0:k}^{k+1:d} \\
    4\alpha\hat{A}_{0:k}^{k+1:d} & I
    \end{pmatrix}
\end{align*}
 Noting that \(\hat{a}^\dagger-\hat{a}\) is tridiagonal, the only
nonzero element of the upper and lower off-diagonal blocks is that at
\(\op{k}{k+1}\) and \(\op{k+1}{k}\). Thus, \(V_k(\alpha)\) only acts
non-trivially on the \(2\times 2\) block at \(\ket{k},\ket{k+1}\), and
on that block acts as: 
\[
\begin{aligned}
V_k(\alpha)_{k:k+1}^{k:k+1} &= 
\begin{pmatrix}
1 & -4\alpha\sqrt{k+1} \\
4\alpha\sqrt{k+1} & 1
\end{pmatrix} + \order{\alpha^2} \\
\end{aligned}
\]

Noting that a Givens rotation expands to first order in \(\theta\) as \[
G(\theta) = \begin{pmatrix}
\cos{\theta} & -\sin{\theta} \\
\sin{\theta} & \cos{\theta}
\end{pmatrix} = \begin{pmatrix}
1 & -\theta \\
\theta & 1
\end{pmatrix}+\order{\alpha^2}
\] we can match orders and see that these two expressions for \(G(\theta)\)
and \(V_k(\alpha)\) match if we set \(\alpha=\Phi(\theta)\) where \(\Phi(\theta)\) \[
    \Phi(\theta) = \frac{\theta}{4\sqrt{k+1}}
\]

Thus, for small \(\theta\), \(V_k(\frac{\theta}{4\sqrt{k}}) \approx G(\theta)\) to
first order.

\hypertarget{numerical-tests-exact-sun-decomposition}{%
\section{Numerical tests \& exact SU(N)
decomposition}\label{numerical-tests-exact-sun-decomposition}}

To analyze the performance of this approximation numerically, we define
the infidelity \(\epsilon\):
\[\epsilon = 1 - F = 1 - \frac1d \abs{\Tr(U^\dagger U_{target})}^2\] for
a given implemented unitary \(U\) and target unitary \(U_{target}\).

In Figure 1 
 we plot the infidelity of Givens rotations
between adjacent Fock states for various angles \(\theta\) and target levels
\(k\) (corresponding to an Givens rotation between levels \(\ket{k}\)
and \(\ket{k+1}\)) for \(k=\{0,1,…,60,61\}\) modeled on a
\(64\)-dimensional qudit by truncating \(D(\alpha)\) to the
\(64\)-dimensional space. We note that we are plotting
\(\max{10^{-15},\epsilon}\) so that we can plot on a log scale, as there are a
number of \(0\) values due to finite floating point precision.

\begin{figure}
\centering{
\includegraphics{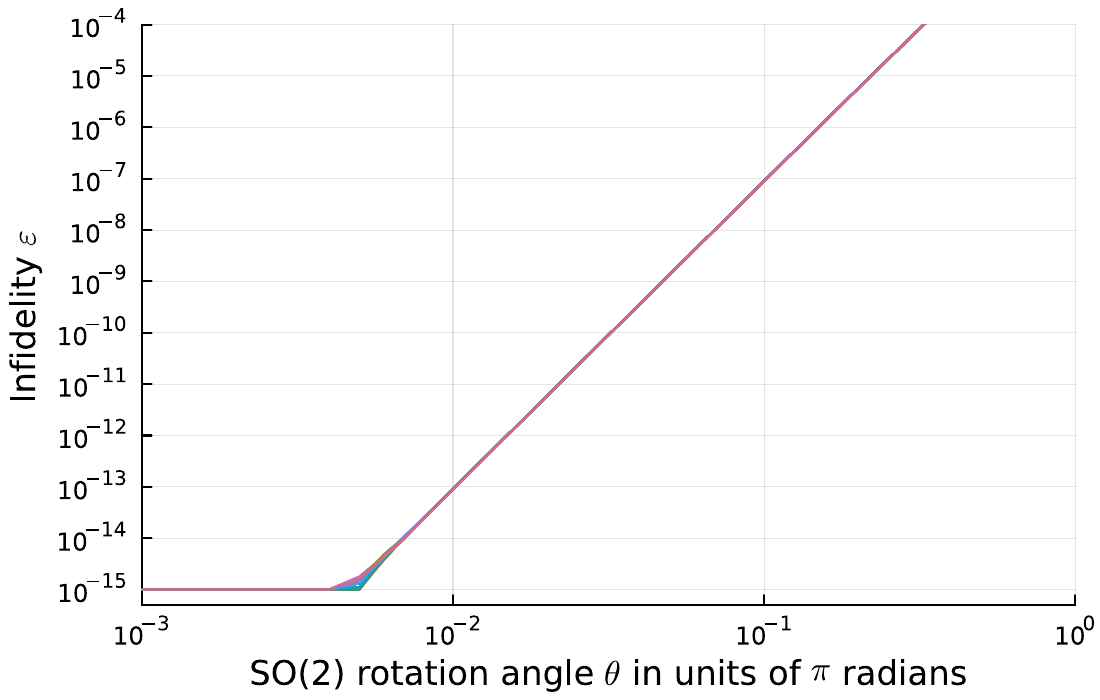}}
\caption{The infidelity \(\epsilon\) of SO(2) operations
(Givens rotations) between levels \(\ket{k}\) and \(\ket{k+1}\) on a
\(64\)-dimensional qudit, implemented via \(V_k(\Phi(\theta))\) as a function of
\(k\), for \(k=\{0,1,…,60,61\}\) (\(k<62\) to avoid significant error
due to truncation of \(D(\alpha)\) to the \(64\) dimensional space).
Different colors denote different values of \(k\), however the curves
line up almost exactly. Values of \(\epsilon<10^{-15}\) are set to \(10^{-15}\)
to avoid floating point accuracy errors yielding \(\epsilon=0\) which cannot be
plotted on a log scale.}
\label{fig:vks02}
\end{figure}

\begin{figure}
\centering{
\includegraphics{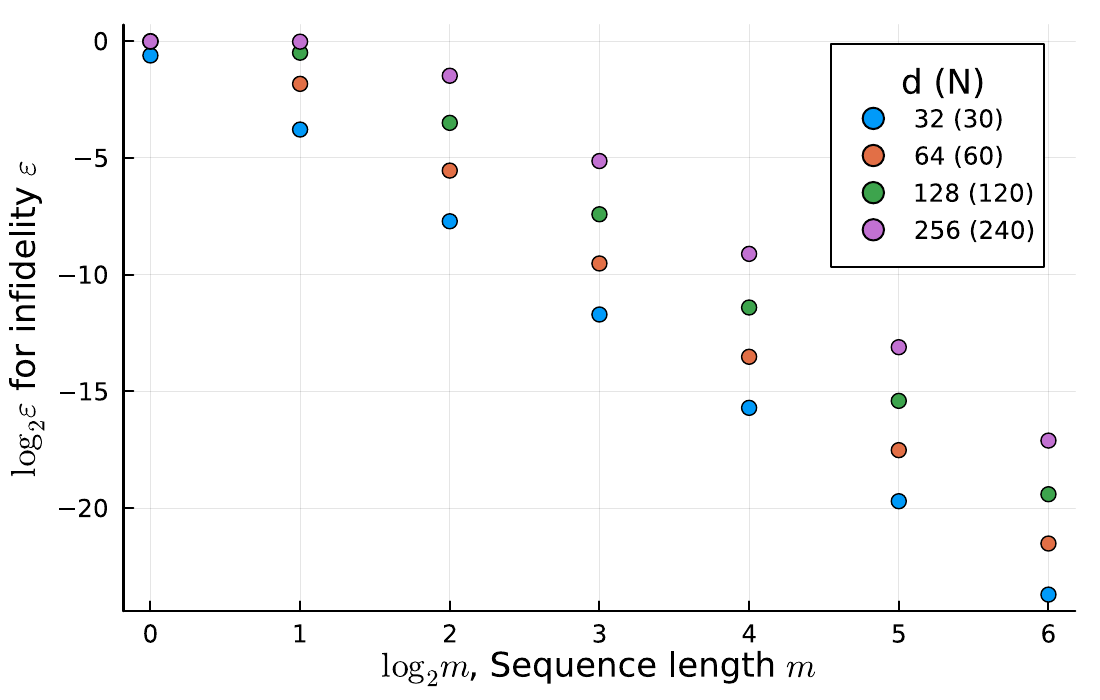}}
\caption{Infidelity of the quantum Fourier transform
gate, compiled via the decomposition into SNAP and Givens rotations,
then via our map \(\alpha=\Phi(\theta)\) into a sequence of SNAP and displacement
operations composed of blocks \(V_k(\alpha)\). We plot infidelity for varying
qudit dimensions (Fourier gate dimension) \(d\) \((N)\) and for a
varying expansion of the compiled gate sequences formed by dividing each
Givens rotation with angle \(\theta\) into \(m\) \(\theta/m\) rotations in order
to improve accuracy. Infidelity scales approximately as
\(\order{m^{-4}}\) and rises as approximately \(\order{d^2}\).}
\label{fig:qft}
\end{figure}

We can see that \(\epsilon \approx \order{\theta^6}\) (a straight line on the
log-log curve between \(\epsilon=\pi/100\) and \(\epsilon=\pi/2\) yields a slope of
\(5.94\)), dramatically better than might be expected since \(\Phi(\theta)\) is
derived via a first order expansion.

To investigate the accuracy of the algorithm for direct unitary
compilation, we compile a quantum Fourier transform gate defined on
\(N\) states as in Ref. ~{[}2{]}: \[
\mathcal{F}_N = \left(\mathcal{F}_N\right)_{l,m} = \frac{1}{\sqrt{N}} e^{i\left[\left(l - N/2\right)\left(m - N/2\right) \right]2\pi/N}
\] and compute the infidelity of that gate as implemented using the
compiled ensemble of SNAP gates and Givens rotation angles \(\vec{\theta}\)
implemented via \(V_k(\Phi(\theta)\) gates.

Further, we note that, since we find that the infidelity of \(V_k(\Phi(\theta)\)
grows as approximately \(\order{\theta^6}\), we can implement a higher
quality, lower error rotation by breaking the Givens rotation into \(m\)
individual \(\theta/m\) rotations for some integer \(m\geq 1\).

We model this for QFT gates \(F_N\) for \(N=\{30,60,120,240\}\) inside
of \(d=\{32,64,128,256\}\)-dimensional qudits, respectively). The
compilation applies to the entire \(d \times d\) unitary, with the
action on the first \(N\) levels being the QFT \(F_N\) and the action of
remaining levels being the identity. \(N<d\) to avoid the displacement
gates in \(V_k\) being significantly affected by truncation to the
finite \(d\)-dimensional space.

Plotting the infidelity \(\epsilon\) against \(m\) and \(d\) in
Figure 2 
, we observe that for all qudit sizes and
sufficiently large \(m\) the infidelity of the entire QFT gate \(F_N\)
falls as approximately \(\epsilon\approx\order{m^{-4}}\) (the slope on a
log-log plot for \(d=256\), for instance, between \(m=8\) and \(m=64\)
is \(-3.991…\)).

Taking that the infidelity for a full \(d \times d\) unitary will scale
as \(\order{d^2}\), as the number of Givens rotations required will
scale as \(\order{d^2}\), then a constant error budget \(\epsilon_{target}\)
can be maintained as \(d\) scales so long as \(m = \order{\sqrt{d}}\).

Therefore, we can decompose any arbitrary unitary into a sequence of
\(\order{d^{5/2}}\) SNAP and displacement gates directly, without
numerical optimization at compile time. As the compilation process
requires a sequence of \(d(d-1)/2\) SO(2) operations to be applied to a
unitary matrix, and applying an SO(2) rotation to a unitary requires
\(4d\) operations, the compilation process uses \(\sim 2d^3\) complex
floating point multiply-add operations, and produces a sequence of SNAPs
and Givens rotations that can, in constant time per rotation, be
translated into a sequence of SNAP and displacement gates. For
reference, on a 2017 13-inch Macbook Pro, this can be done on a single
core in approximately one second for a \(256 \times 256\) dimensional
unitary.

\hypertarget{conclusions}{%
\section{Conclusions}\label{conclusions}}

We present here for the first time, to our knowledge, a direct map \(\Phi\)
from the target Givens rotation angle \(\theta\) to the displacement
magnitude \(\alpha\) used in the gate sequence
\(V_k(\alpha) = D(\alpha)R_\pi(k)D(-2\alpha)R_\pi(k)D(\alpha)\), namely
\(\Phi(\theta) = \frac{\theta}{4\sqrt{k+1}}\). This map is derived to by expanding
\(V_k\) to first order in \(\alpha\) and matching terms with a Givens
rotation expanded to first order, however we find that the infidelity of
the gate sequence scales as \(\epsilon \approx \order{\theta^6}\). Breaking up the
Givens rotations into \(m\) separate, smaller rotations of angle
\(\theta/m\), we observe that full compiled unitary infidelity numerically
scales as \(\order{m^{-4}}\).

Combined, these results imply that qudits can have gates compiled in a
runtime that scales as \(\order{d^3}\) complex floating point operations
into gate sequences with constant error of length \(\order{d^{5/2}}\).
This compilation step can be performed in seconds on a modern laptop for
qudits with hundreds of dimensions, far more quickly than other known
qudit compilation techniques into elementary gates and/or operations.

The chief open questions that remain involve how the compiled gate
sequences are affected by error in the implemented SNAP and displacement
gates, and the real-world runtime scaling of the compiled sequences on
hardware to maintain sufficient gate precision ~{[}13{]}. It is also an
open question if more sophisticated sequences like \(V_k\) can be
constructed to shorten the required gate sequence. Finally, it may be
possible to further compile the SNAP and displacement sequence built
here to shorten the gate sequence and approach the minimum sequence
length of \(\order{d}\). We leave these avenues to future work.

\hypertarget{acknowledgment}{%
\section{Acknowledgment}\label{acknowledgment}}

We also acknowledge the SQMS algorithms team for conversations regarding
this work as it was developed, in particular Gabriel Perdue, Sohaib
Alam, and Barış Özgüler. This material is based upon work supported by
the U.S. Department of Energy, Office of Science, National Quantum
Information Science Research Centers, Superconducting Quantum Materials
and Systems Center (SQMS) under contract number DE-AC02-07CH11359.

\hypertarget{refs}{}
\begin{CSLReferences}{0}{0}
\leavevmode\vadjust pre{\hypertarget{ref-gustafson2021prospects}{}}%
\CSLLeftMargin{{[}1{]} }%
\CSLRightInline{E. J. Gustafson, \emph{Prospects for Simulating a
Qudit-Based Model of (1+ 1) d Scalar QED}, Physical Review D
\textbf{103}, 114505 (2021).}

\leavevmode\vadjust pre{\hypertarget{ref-kurkcuoglu2021quantum}{}}%
\CSLLeftMargin{{[}2{]} }%
\CSLRightInline{D. M. Kurkcuoglu, M. S. Alam, A. C. Li, A. Macridin, and
G. N. Perdue, \emph{Quantum Simulation of \(\phi^4\) Theories in Qudit
Systems}, arXiv Preprint arXiv:2108.13357 (2021).}

\leavevmode\vadjust pre{\hypertarget{ref-rebentrost2019quantum}{}}%
\CSLLeftMargin{{[}3{]} }%
\CSLRightInline{P. Rebentrost, M. Schuld, L. Wossnig, F. Petruccione,
and S. Lloyd, \emph{Quantum Gradient Descent and Newton's Method for
Constrained Polynomial Optimization}, New Journal of Physics
\textbf{21}, 073023 (2019).}

\leavevmode\vadjust pre{\hypertarget{ref-lloyd2020quantum}{}}%
\CSLLeftMargin{{[}4{]} }%
\CSLRightInline{S. Lloyd, M. Schuld, A. Ijaz, J. Izaac, and N. Killoran,
\emph{Quantum Embeddings for Machine Learning}, arXiv Preprint
arXiv:2001.03622 (2020).}

\leavevmode\vadjust pre{\hypertarget{ref-heeres2015cavity}{}}%
\CSLLeftMargin{{[}5{]} }%
\CSLRightInline{R. W. Heeres, B. Vlastakis, E. Holland, S. Krastanov, V.
V. Albert, L. Frunzio, L. Jiang, and R. J. Schoelkopf, \emph{Cavity
State Manipulation Using Photon-Number Selective Phase Gates}, Physical
Review Letters \textbf{115}, 137002 (2015).}

\leavevmode\vadjust pre{\hypertarget{ref-krastanov2015universal}{}}%
\CSLLeftMargin{{[}6{]} }%
\CSLRightInline{S. Krastanov, V. V. Albert, C. Shen, C.-L. Zou, R. W.
Heeres, B. Vlastakis, R. J. Schoelkopf, and L. Jiang, \emph{Universal
Control of an Oscillator with Dispersive Coupling to a Qubit}, Physical
Review A \textbf{92}, 040303 (2015).}

\leavevmode\vadjust pre{\hypertarget{ref-fosel2020efficient}{}}%
\CSLLeftMargin{{[}7{]} }%
\CSLRightInline{T. Fösel, S. Krastanov, F. Marquardt, and L. Jiang,
\emph{Efficient Cavity Control with SNAP Gates}, arXiv Preprint
arXiv:2004.14256 (2020).}

\leavevmode\vadjust pre{\hypertarget{ref-khaneja2005optimal}{}}%
\CSLLeftMargin{{[}8{]} }%
\CSLRightInline{N. Khaneja, T. Reiss, C. Kehlet, T. Schulte-Herbrüggen,
and S. J. Glaser, \emph{Optimal Control of Coupled Spin Dynamics: Design
of NMR Pulse Sequences by Gradient Ascent Algorithms}, Journal of
Magnetic Resonance \textbf{172}, 296 (2005).}

\leavevmode\vadjust pre{\hypertarget{ref-petersson2020discrete}{}}%
\CSLLeftMargin{{[}9{]} }%
\CSLRightInline{N. A. Petersson, F. M. Garcia, A. E. Copeland, Y. L.
Rydin, and J. L. DuBois, \emph{Discrete Adjoints for Accurate Numerical
Optimization with Application to Quantum Control}, arXiv Preprint
arXiv:2001.01013 (2020).}

\leavevmode\vadjust pre{\hypertarget{ref-petersson2021optimal}{}}%
\CSLLeftMargin{{[}10{]} }%
\CSLRightInline{N. A. Petersson and F. Garcia, \emph{Optimal Control of
Closed Quantum Systems via b-Splines with Carrier Waves}, arXiv Preprint
arXiv:2106.14310 (2021).}

\leavevmode\vadjust pre{\hypertarget{ref-Juqbox_osti_1770280}{}}%
\CSLLeftMargin{{[}11{]} }%
\CSLRightInline{N. A. Petersson, F. Garcia, and U. N. N. S.
Administration,
\emph{\href{https://doi.org/10.11578/dc.20210310.3}{Juqbox.jl}},
(2021).}

\leavevmode\vadjust pre{\hypertarget{ref-machnes2018tunable}{}}%
\CSLLeftMargin{{[}12{]} }%
\CSLRightInline{S. Machnes, E. Assémat, D. Tannor, and F. K. Wilhelm,
\emph{Tunable, Flexible, and Efficient Optimization of Control Pulses
for Practical Qubits}, Physical Review Letters \textbf{120}, 150401
(2018).}

\leavevmode\vadjust pre{\hypertarget{ref-alam2022quantum}{}}%
\CSLLeftMargin{{[}13{]} }%
\CSLRightInline{M. S. Alam et al., \emph{Quantum Computing Hardware for
HEP Algorithms and Sensing}, arXiv Preprint arXiv:2204.08605 (2022).}

\end{CSLReferences}

\end{document}